\documentclass{iau}
\usepackage{graphicx,natbib}

\title[Jets in Sources Straddling the Fanaroff-Riley Divide] {Understanding Jets from Sources Straddling the Fanaroff-Riley Divide}

\author[P. Kharb et al.] {Preeti Kharb$^1$, Ethan Stanley$^2$, Matthew Lister$^2$, Herman Marshall$^3$, Chris O'Dea$^{4, 5}$ \and Stefi Baum$^{4, 5}$}
\affiliation{$^1$Indian Institute of Astrophysics, II Block, Koramangala, Bangalore 560034, India \\ email: {\tt kharb@iiap.res.in} \\[\affilskip]
$^2$Dept. of Physics and Astronomy, 525 Northwestern Avenue, West Lafayette, IN 47907, USA \\
$^3$MIT Kavli Institute for Astrophysics and Space Research, 77 Massachusetts Avenue, Cambridge, MA 02139, USA\\
$^4$University of Manitoba, Winnipeg, MB R3T 2N2 Canada\\
$^5$Rochester Institute of Technology, 84 Lomb Memorial Drive, Rochester, NY 14623, USA}

\pubyear{2015}
\volume{313}  
\pagerange{xxx -- xxx}
\jname{Extragalactic Jets from every angle}
\editors{F. Massaro, C. C. Cheung, E. Lopez, and A. Siemiginowska, eds.}

\def\arcsec{^{\prime\prime}}

\begin{document}
\maketitle
\begin{abstract}
Results from {\it Chandra-HST-VLA} observations of 13 hybrid sources are presented. Data from ten sources in the literature are analysed along with new data from three hybrid blazars belonging to the MOJAVE sample. Studies of such hybrid sources displaying both FRI and FRII jet characteristics could provide the key to resolving the long-standing Fanaroff-Riley dichotomy issue. A majority of the 13 hybrid sources show FRII-like total radio powers, i.e., they are ``hybrid'' in radio morphology but not in total radio power. VLBI observations of ten of the 13 sources show that the X-ray jet is on the same side as the one-sided VLBI jet. X-rays are therefore emitted from relativistically-boosted approaching jets. This is consistent with the X-ray emission being IC/CMB in origin in the majority of sources. It is not completely clear from our study that asymmetries in the surrounding medium can create hybrid sources. Hybrid radio morphologies could also be indicative of intrinsically asymmetric jets.
\end{abstract}

\firstsection 
\section{Background}
The Fanaroff-Riley (FR) dichotomy has been a long-standing open question in active galactic nuclei (AGN) astrophysics. It was in 1974 that B. Fanaroff and J. Riley pointed out that sources which possessed radio jets spanning tens to hundreds of kiloparsecs seemed to exhibit primarily two radio morphologies: the FR type I sources had broad jets that flared on scales of a kiloparsec with the brightest emission coming from near the radio cores and fading away with distance into diffuse radio plumes/lobes, while the FR type II sources had collimated jets that terminated in regions of high surface brightness called ``hot spots" with the back-flowing plasma or plasma left behind by the advancing jet forming the radio lobes \citep{Fanaroff74,Bridle84}. FRIs were therefore termed ``edge-darkened''  while FRIIs were ``edge-brightened'' (see Figure 1). The total radio power of the FRI/FRII sources also differed, with the cleanest separation occurring at low radio frequencies: the FRI sources had radio luminosities at 178~MHz below $\approx2\times10^{25}$~W~Hz$^{-1}$, while the FRII sources had 178~MHz luminosities above this value. It was shown in the 1990s that the FRI/FRII division was also a function of the optical luminosity of the host galaxy \citep{Ledlow96}. In the so-called Owen-Ledlow plot, there is clearly a region in the radio luminosity space where FRIs and FRIIs can co-exist \citep[see also][]{Baum95}: this is typically between Log~L$_{1.4\,\mathrm{GHz}}$ of $\approx24.5-26.0$~W~Hz$^{-1}$. The relevance of this will be discussed ahead in Section~2.1.

It was soon realised that many different types of AGN were the same phenomena modified in appearance due to orientation-effects of relativistic jets and the presence of dusty tori which shielded the central regions containing the black hole, accretion disk and the broad line region, from certain lines of sight. This gave rise to the radio-loud Unified Scheme which postulated that BL~Lac objects were the pole-on counterparts of FRI radio galaxies and radio-loud quasars were the pole-on counterparts of  FRII radio galaxies \citep{Urry95}. BL~Lac objects and radio-loud quasars are collectively referred to as blazars. 

However, several burning questions remain. Why do only $\sim15\%-20\%$ of AGN \citep[e.g.,][]{Kellermann89} produce the large radio jets observed in radio galaxies and blazars~? Why do they exhibit distinctly different radio morphologies and possess characteristic radio powers~? How do these radio jets interact with the surrounding matter in the interstellar/intergalactic medium~? 

It was noticed early on, however, that not all radio galaxies fit neatly into the FRI/FRII categories \citep[see examples of sources with complex radio morphologies in][]{Bridle94}. Some sources seem to have one jet that terminates in a terminal hot spot like an FRII, while the other jet fades in surface brightness with distance from the radio core without forming a terminal hotspot, like an FRI \citep[e.g.,][Figure~1]{Morganti93}. These Intermediate or Hybrid sources straddling the Fanaroff-Riley divide are the focus of the present work.

\subsection{The Importance of being Intermediate}
Several suggestions have been put forward to explain the FR dichotomy. These include suggested differences in black hole masses with FRIs possessing higher mass black holes than FRIIs \citep[e.g.,][]{Ghisellini01}; differences in the spin rates of black holes with FRIIs having faster spinning black holes \citep[e.g.,][]{Meier99}; differences in the accretion rates with FRIIs typically having higher accretion rates \citep[e.g.,][]{Marchesini04}, and differences in jet composition with FRI jets comprising electrons and positrons and FRII jets comprising electrons and protons \citep[e.g.,][]{Celotti93}. Such differences cannot account for the existence of Hybrid or Intermediate sources. Although for the last suggestion of jet composition, a variant of an electron-positron ``spine'' surrounded by an electron-proton ``sheath'' could still work for jets in hybrid sources \citep[e.g.,][]{Pelletier89}. As such, the study of hybrid sources could go a long way in helping us resolve the Fanaroff-Riley dichotomy.

\subsection{How Rare are Hybrid Sources ?}
After parsing through the radio images of nearly a thousand sources in the literature, \citet{GopalKrishna00} identified only six sources exhibiting a hybrid radio morphology. The detection rate for such sources was therefore quoted at $<1\%$. \citet{Gawronski06} studied the VLA FIRST survey sources and after creating a sub-sample of 1700 sources with 1.4~GHz flux densities $S_{1.4} >20$ mJy and radio extents $>8\arcsec$, identified five hybrid sources. The detection rate therefore remained at $<1\%$. 

We searched for hybrid morphology sources in the complete flux density limited MOJAVE\footnote{Monitoring Of Jets in AGN with VLBA Experiments} sample of blazars \citep{Lister09}. The parsec-scale jets in the MOJAVE blazars have been regularly monitored with Very Long Baseline Interferometry (VLBI) at 15~GHz for nearly two decades now. This has aided in the determination of jet speeds on parsec-scales for individual jet components/knots over a long time baseline. \citet{Kharb10} observed the MOJAVE-1 sample\footnote{http://www.physics.purdue.edu/astro/MOJAVE/MOJAVEtable.html} with the VLA A-array configuration at 1.4 GHz. Through this study, they identified a much larger fraction of hybrid sources at $\approx8\%$. This larger fraction is likely to be the result of the MOJAVE selection criteria, which are based on relativistically boosted parsec-scale jet emission, rather than low frequency lobe emission like the 3C sample. Therefore, the MOJAVE survey may be picking up sources that span a larger range in intrinsic radio powers and morphologies. As discussed in \citet{Kharb10}, there may also be a greater tendency to pick up jets that are highly bent. 

\begin{figure}[t]
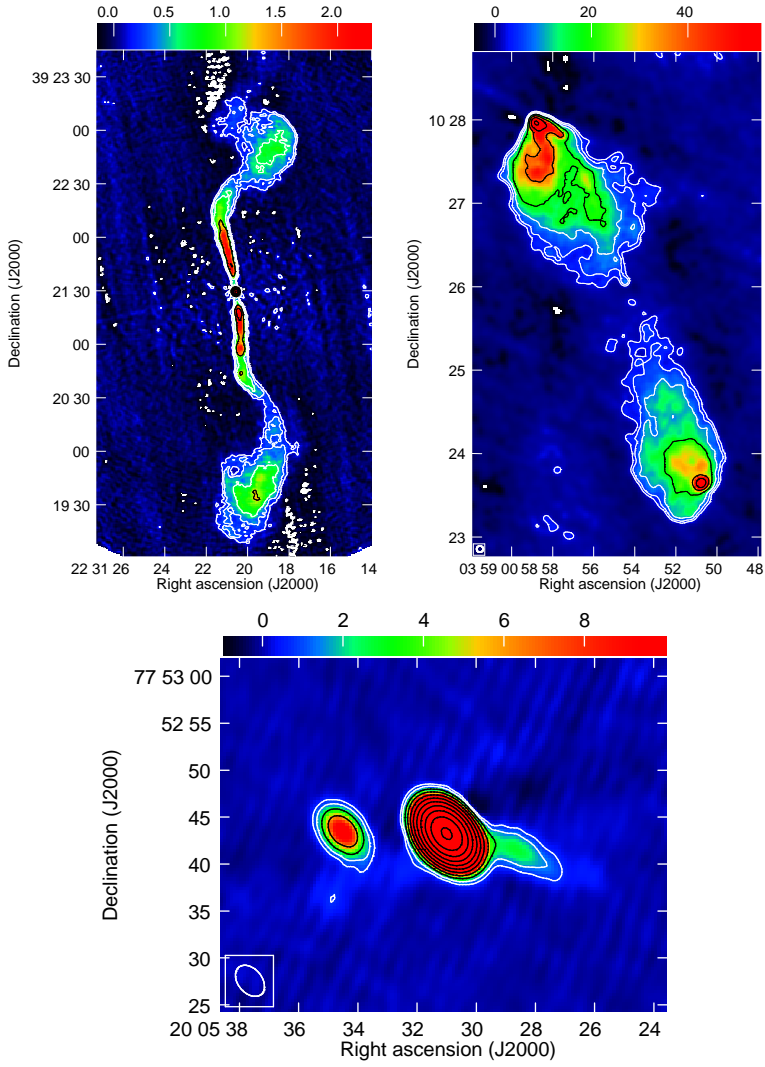

\centering{
\includegraphics[height=8cm]{Kharb_f1a.ps}
\includegraphics[height=8cm]{Kharb_f1b.ps}
\includegraphics[width=8cm]{Kharb_f1c.ps}}
\caption{Radio images of the FRI galaxy 3C\,449 (Top left) and the FRII galaxy 3C\,98 (Top right). The brightest radio emission, seen in red, is closest to the radio core in FRIs and furthest away from the core, at the termination points of the jets, in FRIIs. (Bottom) The radio image of the hybrid morphology blazar, 2007+777, clearly shows an FRI-like jet on the right side and an FRII-like jet (actually only the terminal hot spot) on the left side of the radio core. In this source the one-sided VLBI jet is pointing along the FRI (i.e., the right) side.}
\end{figure}

\subsection{Previous Chandra$-$HST$-$VLA Observations of FRIs \& FRIIs}
The {\it Chandra X-ray Observatory} has demonstrated clearly that X-ray jets are prevalent in blazars \citep{Sambruna04,Marshall05,Hogan11}. A general trend has been observed in the X-ray emission mechanisms of FRI and FRII sources, with the former typically emitting via the synchrotron mechanism \citep[e.g.,][]{Sambruna07,Kharb12b} and the latter emitting via the inverse-Compton scattering of CMB photons \citep[IC/CMB, e.g.,][]{Tavecchio00,Kharb12a}. However, as highlighted in the review by \citet{Harris06}, each of the suggested emission mechanisms are fraught with difficulties. While it is impossible to get synchrotron X-ray emission from 100 kpc scale jets without invoking {\it in situ} particle acceleration throughout the extent of the jets on account of the short lifetimes of X-ray emitting electrons ($\sim$few years), the requirement that the kpc-scale jets stay highly relativistic with bulk Lorentz factors, $\Gamma\sim5-10$, is contradicted by the $\Gamma\sim2$ implied from radio observations of kpc-scale jets \citep{Bridle94,Mullin09}. The Synchrotron Self-Compton (SSC) mechanism typically requires the jet magnetic field strengths to be well below the ``equipartition'' values, for producing X-rays from jets.

\subsection{Previous Chandra$-$HST$-$VLA Observations of Hybrid Sources}
Ten hybrid morphology sources have so far been observed with the {\it Chandra} and in some cases also the {\it Hubble Space Telescope (HST)}. The results from the multi-band SED modelling of these sources are mixed. (Note that we have not considered {\it bona-fide} X-shaped radio sources that consist of two pairs of radio lobes, one pair possessing terminal hot spots and the other pair not, into the hybrid category. However, a couple of sources with ambiguous morphologies that have been classified as X-shaped sources, like 3C\,433 or NGC\,6251, are included in our analysis.) The multi-wavelength observations of hybrid sources have shown that the X-ray jets are sometimes synchrotron in origin \citep[e.g.,][]{Sambruna07}, but more often from the IC/CMB mechanism \citep[e.g.,][]{Miller06,Miller09}.

\subsection{New Multi wavelength Observations of Three MOJAVE Hybrid Blazars}
In order to augment the multi-wavelength data on hybrid morphology sources in the literature, as well as to make greater sense of the results obtained from only a handful of sources so far, we observed three MOJAVE blazars exhibiting a hybrid morphology with the {\it Chandra} ($\sim60-80$ ks using the {ACIS}) and {\it HST} (2 orbits each using the {WFC3/F160W} and {F475W} filters). The first results from this study will be presented in E. Stanley et al. 2015 (in preparation). The three blazars {\it viz.,} 1045$-$188, 1849+670, 2216$-$038, have 1.4~GHz flux densities $>100$~mJy and radio jet extents $>3\arcsec$. These three have been observed previously for $\sim$10 ks each with the {\it Chandra} as part of the MOJAVE-Chandra Sample (MCS) pilot study \citep{Hogan11}.

\section{Results and Comparison with Sources in the Literature}
Multi-waveband SED modelling indicates that the X-ray emission is from the IC/CMB mechanism in all three MOJAVE hybrid blazars that we studied. In addition, the SED best-fit parameters are similar to those observed in regular (non-hybrid) blazar jets: Doppler factors of $\sim3-4$, magnetic fields of $\sim50\mu$G, minimum electron Lorentz factors of $\sim10-20$, maximum electron Lorentz factors of $\sim10^5-10^6$ and electron power law indices of $\sim3-4$ (Stanley et al. 2015). Considering the ten hybrid sources in the literature along with the three MOJAVE blazars, we observe some trends. For 11 of the 13 sources, the X-ray jet is on the side without a terminal hot spot, i.e., the FRI side. For only two of the 13 sources, viz., 3C\,371, 1045$-$188, is the X-ray jet on the side with the terminal hot spot, the FRII side. Finally, for the 11 sources with X-rays on the FRI side, eight times ($>70\%$) the emission mechanism is consistent with the IC/CMB mechanism, while only in three cases the emission is likely to be synchrotron. For the two sources with the X-rays on the FRII side, in one case the X-ray emission is from the IC/CMB and in another the synchrotron mechanism.

\begin{figure}[t]
\centerline{
\includegraphics[width=7.5cm,trim=30 5 25 0]{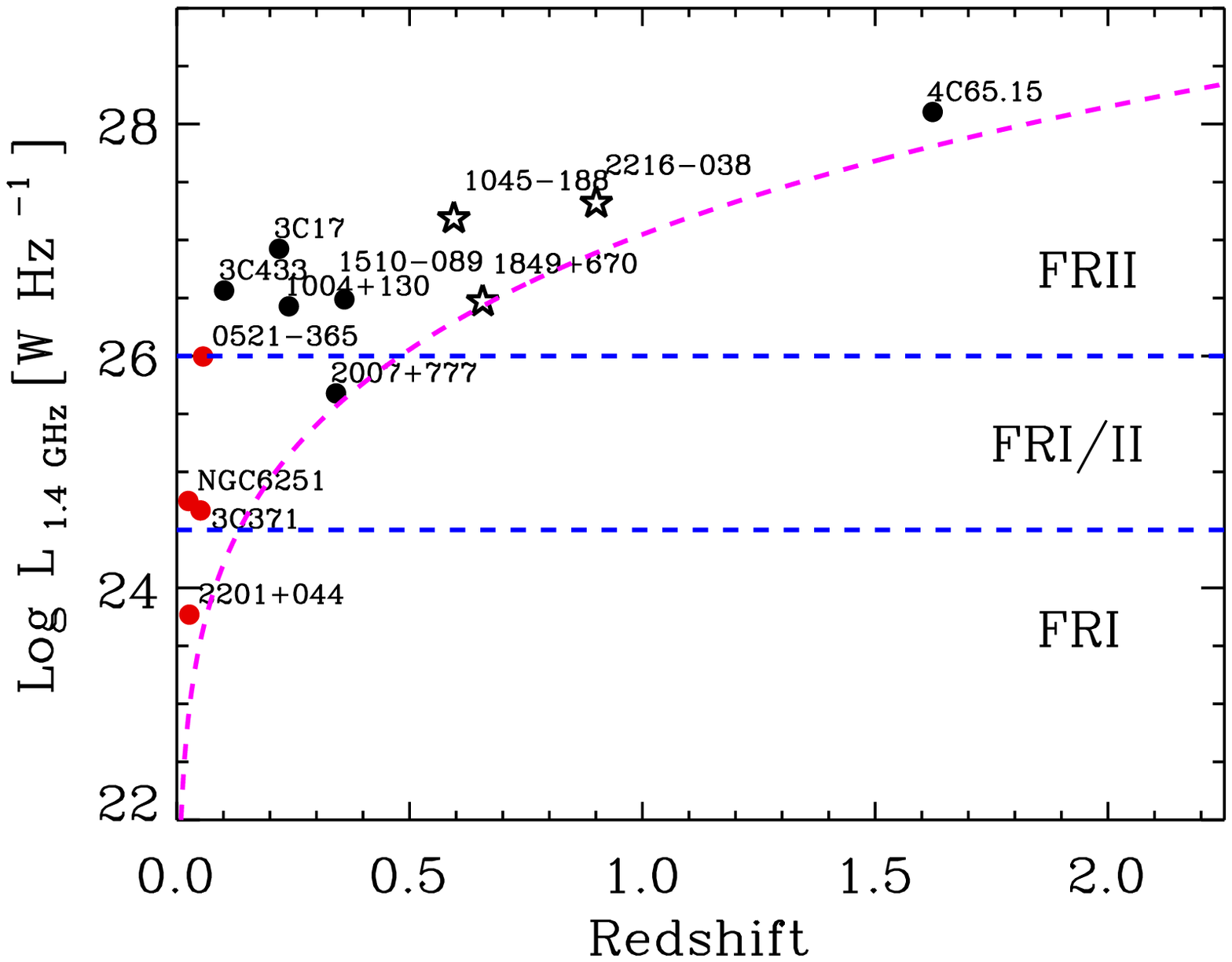}
\includegraphics[width=7.5cm,trim=30 5 25 0]{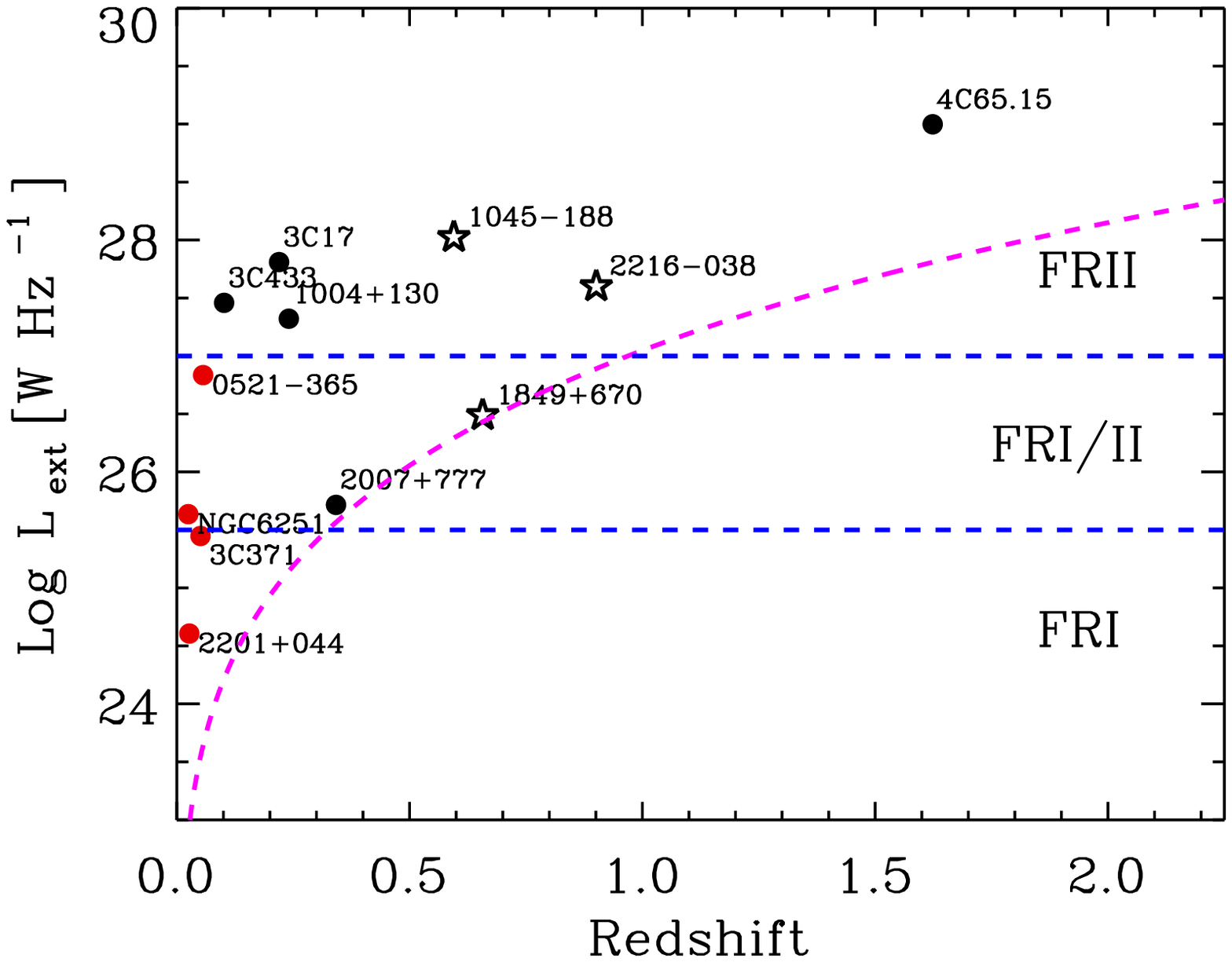}}
\caption{Total radio luminosity at 1.4~GHz (left) and extended radio luminosity at 74~MHz (right) w.r.t. redshift. The red symbols denote sources with X-ray jets from the synchrotron mechanism ($z<0.056$) while the black symbols denote sources with IC/CMB X-ray jets. The open stars denote the three MOJAVE blazars. The magenta line indicates the detection limit of the VLSS.}
\end{figure}

\subsection{``Hybrid'' in Radio Morphology But Not in Radio Power}
We looked at the total and extended (=total$-$core) emission on kpc-scales at 1.4~GHz and 74~MHz respectively for the 13 hybrid sources. The total 1.4~GHz flux densities were derived from the total 74~MHz flux densities assuming a spectral index $\alpha$ of 0.7 ($S_\nu\propto\nu^{-\alpha}$). The 74~MHz flux density values were obtained from the VLA Low Frequency Survey \citep[VLSS;][]{Cohen07}. For estimating the extended flux densities we obtained the core flux densities from high resolution 8~GHz images in the NRAO VLA Image archive\footnote{https://archive.nrao.edu/archive/archiveimage.html} and the Gaussian-fitting AIPS task {\tt JMFIT}, and subtracted them from the total 74~MHz flux densities. We note that if our assumption that the core spectral indices are flat ($\alpha=0$) all the way down to 74~MHz is incorrect, and the spectrum turns over at low frequencies, then our estimated extended flux densities are in fact lower limits. The total and extended luminosities for the hybrid sources are plotted against redshift in Figure~2. Blue dashed lines indicate the FRI, FRII and intermediate FRI/II categories (see Section 1). In the right panel of Figure~2, we have shifted the FR dividing lines assuming $\alpha=0.7$. The magenta line indicates the detection limit of the VLSS. The red symbols denote sources with X-ray jets from the synchrotron mechanism (all have redshifts $<0.056$) while the black symbols denote sources with IC/CMB X-ray jets. As is clear from these figures, most of the hybrid sources, while ``hybrid'' in radio morphology, are not hybrid in radio power. Rather they have FRII-like total radio powers. It becomes apparent from Figure~2 that the total radio power, rather than the FR radio morphology, is the main determinant for the X-rays being IC/CMB (in high power sources) or synchrotron in origin (in low power sources).

\subsection{Information from Parsec-scales}
Ten of the 13 hybrid sources have so far been imaged with VLBI. However, only seven of them have been imaged at multiple epochs with VLBI to yield apparent jet speeds. We find that the one-sided VLBI jets are always on the same side as X-ray jets except apparently in the blazar 1510$-$089. However, \citet{Homan02} have shown that 1510$-$089 is a blazar with a highly bent jet oriented at a small angle to line of sight. Therefore, it appears that in all cases so far, the X-ray emission is from the relativistically-boosted approaching jet. 

\begin{figure}[t]
\centerline{
\includegraphics[width=7.5cm,trim=30 5 25 0]{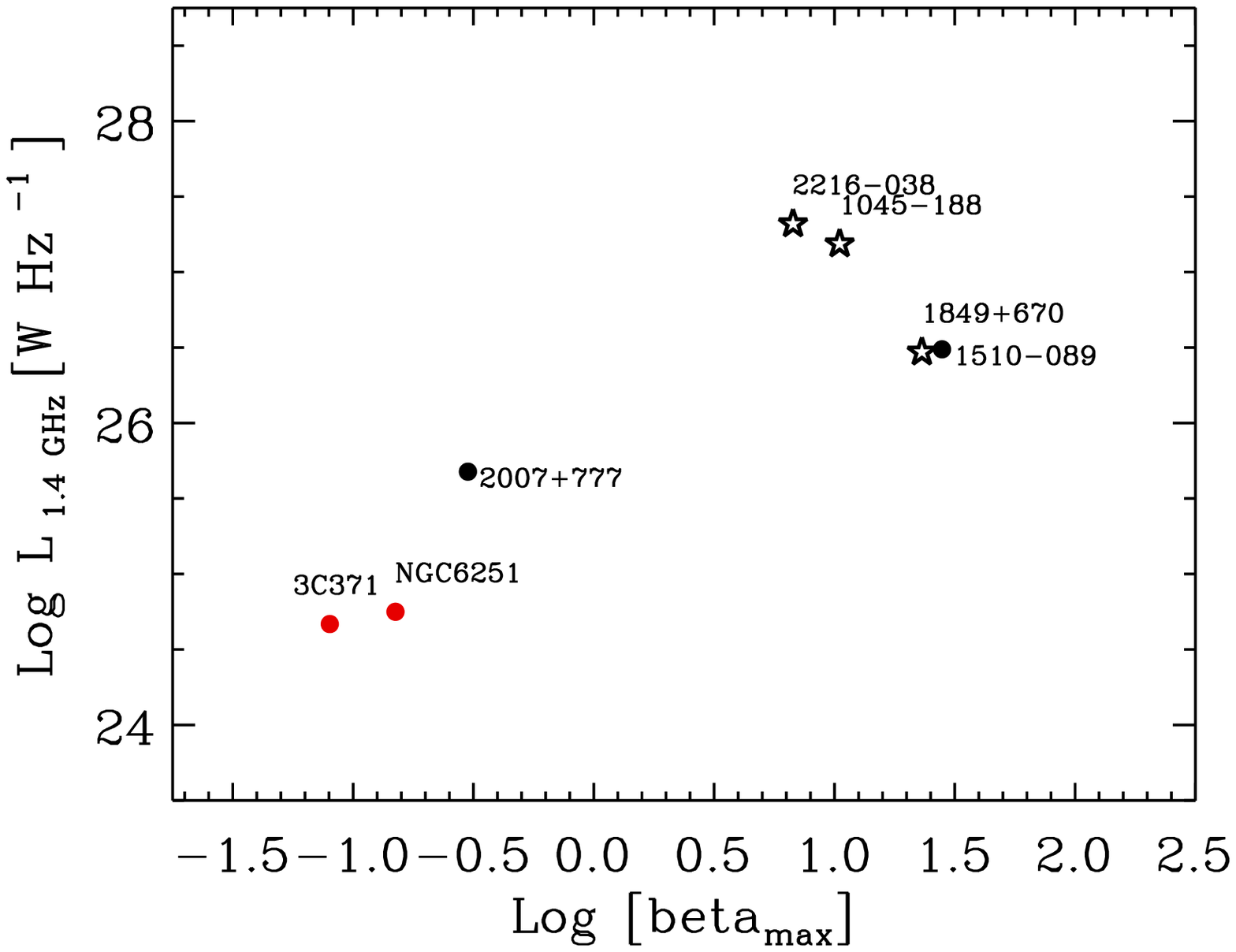}
\includegraphics[width=7.5cm,trim=30 5 25 0]{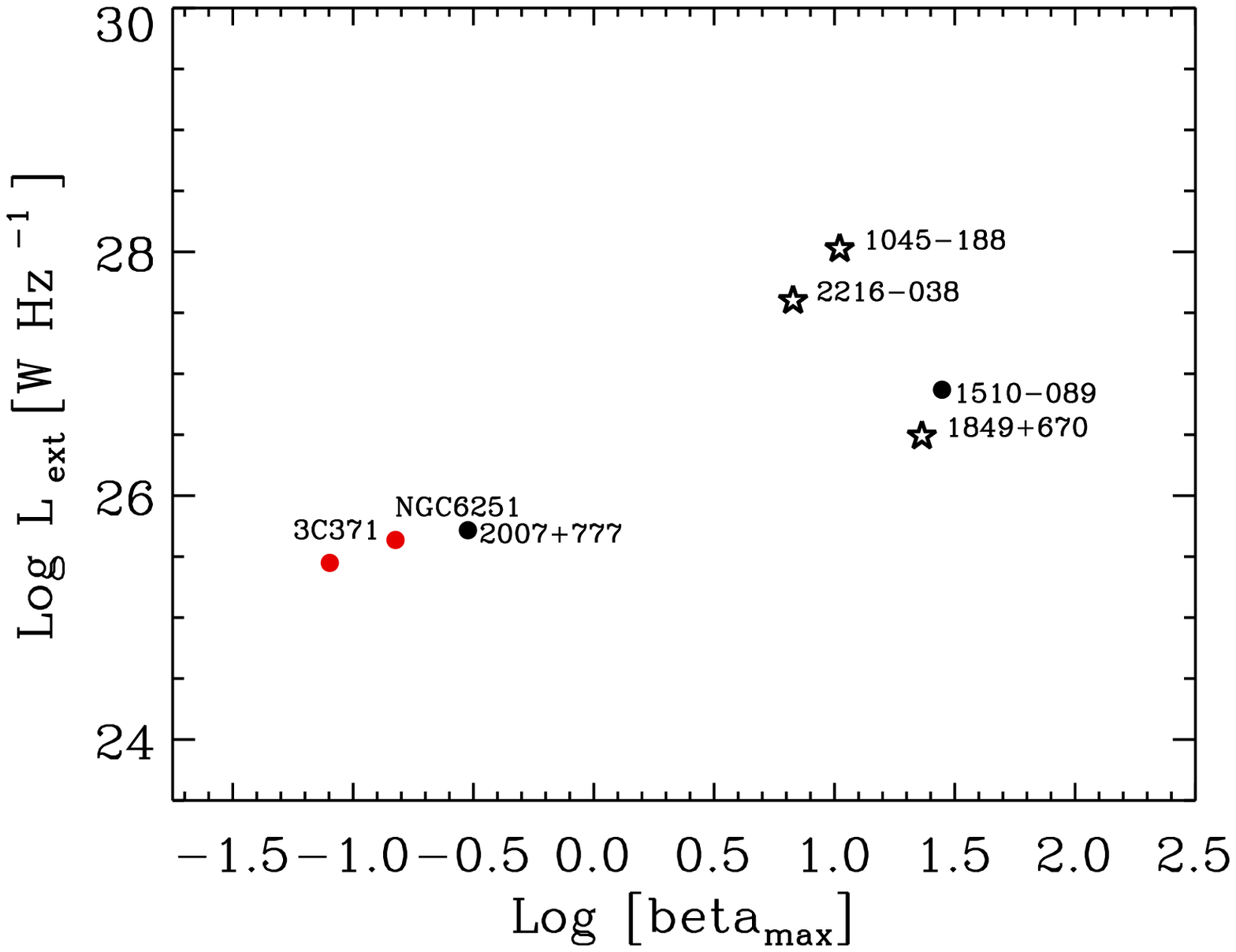}}
\caption{Total (left) and extended (right) radio power versus the maximum apparent jet speeds on parsec-scales. The red symbols denote sources with X-ray jets from the synchrotron mechanism ($z<0.056$) while the black symbols denote sources with IC/CMB X-ray jets. The open stars denote the three MOJAVE blazars.}
\end{figure}

We have examined the total ($L_{1.4\,\mathrm{GHz}}$) and extended luminosities ($L_{ext}$) versus the maximum jet component speeds $\beta_{max}$ for seven hybrid sources in Figure~3.
\cite{Kharb10} had found a significant correlation between $\beta_{max}$, and extended radio luminosity. On average, the radio-loud quasars had faster jets than BL Lac objects. However, no clear division in $\beta_{max}$ or $L_{ext}$ was observed between quasars and BL Lacs. The $\beta_{max}-L_{1.4\,\mathrm{GHz}}$ and $\beta_{max}-L_{ext}$ correlation is also observed for the seven hybrid sources with measured VLBI jet speeds (Figure 3). We see that the two sources with synchrotron X-ray emission (red symbols) are at the lower end of the $\beta_{max}-$luminosity correlations, while the X-ray jets from the IC/CMB process are towards the higher end. Also for the few sources considered here the BL Lac objects (NGC\,6251 is an FRI radio galaxy) and quasars are clearly divided below and above a $\beta_{max}$ of around 3. This study will benefit from multi-epoch VLBI observations of a greater number of hybrid sources.

\section{Are Hybrid Sources FRIIs in Asymmetric Environments ?}
\citet{Miller09} suggested that the hybrid morphology sources were likely to be FRIIs in asymmetric environments, consistent with the suggestions of \citet{GopalKrishna00}: the jets that traversed the denser media eventually became de-collimated and FRI-like. However, \citet{Miller09} concluded this on the basis of the bent radio jets, rather than any significant asymmetry in the diffuse X-ray emission from and near the host galaxies. The SDSS image of one of the two sources considered in their study did not reveal any over-density of nearby galaxies, while the other source was part of a galaxy group. However, we know that bent radio jets can result from other phenomena like jet nozzle precessions \citep[e.g.,][]{Lister03}. Therefore, the mere presence of bent jets cannot be used to infer strong jet-medium interactions which can change the trajectory of the radio jets.

For the three MOJAVE blazars, a first-order analysis on the diffuse host galaxy X-ray emission shows no asymmetry (Stanley et al. 2015). We next looked at the SDSS images of the 13 hybrid sources and found that eight of them have nearby galaxies, with projected distances $<40-50$ kpc. A few of these sources are well-documented galaxy pairs and some show signatures of galaxy-galaxy interactions. Therefore, $\gtrsim60\%$ of the hybrid sources considered here may indeed have asymmetric galactic environments on account of galaxy interactions. However, to confirm their group or cluster membership, we need the as yet unavailable redshifts of the surrounding galaxies. Results from a more detailed analysis including a comparison sample of regular active galaxies will be presented in Stanley et al. (2015).

However, there is some circumstantial evidence that does not favour the asymmetric environment argument. We know that the X-ray jets are always on the same side as the one-sided VLBI jets. This implies that the X-ray jets are also the approaching jets. Secondly, the majority of the jets that emit in X-rays are on the side without a terminal hot spot, or the FRI side. The asymmetric environment argument predicts that this is the side with the denser medium. However, it would be peculiar if the denser medium was always on the same side as the approaching jet. 

Another question then arises: why is the X-ray emission typically on the side without the terminal hot spot ? This could happen due to unfavourable Doppler boosting effects. For example, for an approaching jet, the backflow from the hot spot could be moving away from us resulting in net surface brightness dimming. This result could also imply that even on the FRI side there is a fast collimated inner spine, just like an FRII jet, but which dissipates before a terminal hot spot is formed.
 
\section{Summary}
Studying sources with morphologies that are intermediate between FRIs and FRIIs could be the key to understanding the Fanaroff-Riley dichotomy. Intrinsic differences in the central engine (e.g., black hole mass, spin, accretion rate, jet composition) can be ruled out for hybrid sources. These studies can in turn provide invaluable clues about jet formation and propagation. While most theories of hybrid sources have invoked asymmetry in the external environment, \citet{Wang92} have proposed that if the magnetic field structure in the accretion disk is asymmetric with respect to the disk mid-plane, then intrinsically asymmetric jets could be launched. Such a model could explain the presence of hybrid morphology sources. We need to further explore this and similar models.

We summarise below the results from the {\it Chandra-HST-VLA} observations of 13 hybrid sources. Data from ten sources in the literature are analysed along with new data from three hybrid blazars belonging to  the MOJAVE sample.
\begin{enumerate}
\item A majority of the 13 hybrid sources considered in this study show FRII-like total radio powers. Therefore, these hybrid sources are ``hybrid'' only in radio morphology but not in total radio power.
\item For ten of the 13 hybrid sources with parsec-scale data from VLBI, the one-sided VLBI jets which are the relativistically-boosted approaching jets, are on the same side as the X-ray emission. Therefore, the X-rays are emitted from approaching jets. This is consistent with the X-ray emission being IC/CMB in origin, in the majority of sources.
\item It appears that the total radio power, rather than the FR radio morphology, is the main determinant for the X-rays being IC/CMB (in high power sources) or synchrotron in origin (in low power sources).
\item There is a tendency for the X-ray jets to lie on the side without the terminal hot spots, i.e., the FRI side. This could occur due to unfavourable Doppler boosting effects: for an approaching jet, the hot spot backflow could be moving away from us resulting in surface brightness dimming. 
\item It is not completely clear that asymmetries in the surrounding medium can create hybrid sources from regular FRII sources. It is expected in this scenario that the jet traversing the denser medium would get de-collimated and become FRI-like.  While the SDSS images do reveal the presence of nearby ($<40-50$~kpc) galaxies for $\gtrsim60\%$ of the hybrid sources, confirmation requires the yet unavailable redshifts of the apparently nearby galaxies. However, following points ({\it b}) and ({\it d}) above, it would be peculiar to have the denser environment always on the same side as the approaching jets. We need to further explore the realm of intrinsically asymmetric jets to produce hybrid radio morphologies. 
\end{enumerate}
 
\section*{Acknowledgements}
PK would like to thank the organisers for the invitation to speak at this excellent conference.
This work was supported by the National Aeronautics and Space Administration (NASA) through Chandra Award Number (GO3-14120A) issued by the Chandra X-ray Observatory Center (CXC), which is operated by the Smithsonian Astrophysical Observatory (SAO) for and on behalf of NASA under contract NAS8-03060. Support for program number 13116 was provided by NASA through a grant from the Space Telescope Science Institute, which is operated by the Association of Universities for Research in Astronomy, Inc., under NASA contract NAS5-26555. The MOJAVE program is supported under NASA-Fermi grant NNX12A087G. The National Radio Astronomy Observatory is a facility of the National Science Foundation operated under cooperative agreement by Associated Universities, Inc.
 
\bibliographystyle{apj}
\bibliography{Kharb}
\end{document}